# Effects of High-β on Phase-locking Stability and Tunability in Laterally Coupled Lasers

Sizhu Jiang⸸, Suruj S. Deka⸸, Si Hui Pan, Yeshaiahu Fainman, *Fellow, IEEE*

*Abstract*—Phase-locked laser arrays have been extensively investigated in terms of their stability and nonlinear dynamics. Specifically, enhancing the phase-locking stability allows laser arrays to generate high-power and steerable coherent optical beams for a plethora of applications, including remote sensing and optical communications. Compared to other coupling architectures, laterally coupled lasers are especially desirable since they allow for denser integration and simpler fabrication process. Here, we present the theoretical effects of varying the spontaneous emission factor $\beta$, an important parameter for micro- and nanoscale lasers, on the stability conditions of phase-locking for two laterally coupled semiconductor lasers. Through bifurcation analyses, we observe that increasing $\beta$ contributes to the expanding of the in-phase stability region under all scenarios considered, including complex coupling coefficients, varying pump rates, and frequency detuning. Moreover, the effect is more pronounced for $\beta$ approaching 1, thus underlining the significant advantages of implementing nanolasers with intrinsically high $\beta$ in phase-locked laser arrays for high-power generation. We also show that the steady-state phase differences can be widely tuned – up to $\pi$ radians – by asymmetrically pumping high-$\beta$ coupled lasers. This demonstrates the potential of high-$\beta$ nanolasers in building next-generation optical phased arrays requiring wide scanning angles with ultra-high resolution.

*Index Terms*—Laser dynamics, phase locking, nanolaser, semiconductor lasers, laser arrays, bifurcation analysis

## I. Introduction

PHASE-LOCKED laser arrays have been extensively investigated owing to their potential in generating high-power and coherent optical beams valuable for applications such as LiDAR, optical communications and remote sensing [1-3]. Additionally, the ability to tune the phase difference between constituent elements in an on-chip semiconductor laser array is vital for beam forming and steering applications [4,5]. To realize the desired phase offset of the lasers in the arrays, i.e. in-phase operation for high power emission and shifted phase operation for beam forming and scanning, establishing stable phase locking is imperative. However, such stability is challenging to achieve due to multiple factors such as mode competition, distinct time scales of photon and carrier dynamics, complex nonlinear dynamics over a wide range of physical parameters including inter-cavity distances and differences in resonator dimensions of coupled lasers, and most notably, due to the amplitude-phase coupling in semiconductor lasers quantified by the linewidth enhancement factor [6,7]. Despite the challenges, stable in-phase locking has been demonstrated through spatial and spectral mode engineering, including evanescent coupling in topological cavities [8], non-Hermitian coupling in super-symmetry arrays [9], diffractive coupling via Talbot effect [10], global antenna coupling [11], and gain matching [12]. Achieving similar phase synchronization in laterally coupled lasers arranged in close proximity, although difficult, is highly desirable since it involves simpler fabrication procedures and offers denser on-chip integration compared to the other coupling schemes mentioned above. Moreover, the dynamical behavior of laterally coupled lasers can be accurately analyzed and predicted by non-complex mathematical models.

In fact, theoretical analysis of the stability in laterally coupled semiconductor lasers has been widely reported in the literature [13-18]. In these studies, a plethora of dynamical regimes including stable continuous-wave operation, periodic and period-doubling oscillations, chaos, bistability, and chimera states are identified via bifurcation analysis. These dynamical behaviors are obtained by either analytically or numerically solving the coupled rate equations that govern the temporal dynamics of the emitted laser field. The impact of a variety of important parameters such as current injection rate, linewidth enhancement factor, laser size differences, as well as carrier and photon lifetimes, are addressed in these analyses. However, one critical parameter consistently overlooked in the majority of the theoretical works in the literature thus far is the spontaneous emission factor $\beta$. This $\beta$ factor, defined as the fraction of spontaneous emission funneled into the lasing mode

Manuscript received December 13, 2020. This work was supported by the Defense Advanced Research Projects Agency (DARPA) DSO NLM and NAC Programs, the Office of Naval Research (ONR), the National Science Foundation (NSF) grants DMR-1707641, CBET-1704085, NSF ECCS-180789, NSF ECCS-190184, NSF ECCS-2023730, the Army Research Office (ARO), the San Diego Nanotechnology Infrastructure (SDNI) supported by the NSF National Nanotechnology Coordinated Infrastructure (grant ECCS-2025752), the Quantum Materials for Energy Efficient Neuromorphic Computing-an Energy Frontier Research Center funded by the U.S. Department of Energy (DOE) Office of Science, Basic Energy Sciences under award # DE-SC0019273.Advanced Research Projects Agency Energy (LEED: A Lightwave Energy-Efficient Datacenter), and the Cymer Corporation. (*Corresponding author: Sizhu Jiang*)

S. Jiang, S. S. Deka, S. H. Pan, and Y. Fainman are with the Department of Electrical and Computer Engineering, University of California San Diego, La Jolla, CA 92093-0407 USA (email: sij023@ucsd.edu; sdeka@ucsd.edu; h0pan@ucsd.edu; fainman@eng.ucsd.edu).

⸸These two authors contributed equally to the work.

relative to all the other supported modes, is smaller than $10^{-3}$ in typical commercial laser diodes, and thus is reasonably neglected in most bifurcation studies [15-17]. Over the past two decades, however, nanolasers that exhibit intrinsically high and non-negligible $\beta$ values have been demonstrated on various platforms [19-23]. These nanoscale light sources offer unique advantages such as ultracompact footprints, low power consumption and high-speed modulation that make them ideal candidates for dense, on-chip integration [23-26]. A handful of studies so far have reported on the significant impact that spontaneous emission can exert on coupling behavior such as mode switching for coupled photonic crystal nanolasers [10,27] and partial locking for mutually coupled micropillar lasers operating in the few-photon regime [28, 29]. Notably, some previous theoretical investigations have suggested that the larger damping effect induced by higher $\beta$ may help suppress the instability encountered in lateral coupling schemes [14,30]. This hypothesis as well as the rapid advances in nanolaser technology necessitate an in-depth analysis of how large values of $\beta$ can contribute towards stable phase-locking operation.

In this manuscript, we theoretically investigate the effects of varying $\beta$ on the stability of phase locking in two laterally coupled semiconductor lasers through bifurcation analyses. With increasing $\beta$, a corresponding expansion of the stability region is observed when first considering a purely imaginary coupling coefficient, representing a system where the supermodes have identical losses. In order to assess the feasibility of such stability enhancement due to high-$\beta$ in a more practical device, we also consider other important control parameters such as the pump rate and the resonance frequency detuning between the coupled lasers that can result due to fabrication imperfections. We then further extend the study by including complex coupling coefficients in order to better account for realistic scenarios where the supermodes face dissimilar losses. Increases in the phase-locking stability regions driven by increases in $\beta$ are observed across variations of all the control parameters, thereby confirming that the $\beta$-driven stable phase-locking phenomenon is not merely restricted to any specific pump power level or to only negligibly small frequency detuning. Finally, by pumping the two lasers unequally, it can be shown that the steady-state phase difference between the emitters varies as a function of $\beta$, with higher values of $\beta$ resulting in a wider range of relative phase tunability. Therefore, the results presented here emphasize the significance of using high-$\beta$ nanolasers in phase-locked arrays which can demonstrate both high output power density as well as beam forming and steering capabilities depending on the desired application.

This paper is organized into four sections. Section II elaborates on the theoretical model used to perform the numerical simulations. Then in Section III, the effects of varying $\beta$ on the phase-locking stability is discussed in detail. To provide an intuitive understanding of the results in this section, the model is reduced to be as simple as possible initially and then sequentially increased in complexity, one additional parameter at a time. Specifically, equal pumping, an imaginary coupling coefficient and no frequency detuning are assumed while evaluating the effects of altering $\beta$ in Section III-A. Then, the pump rate, frequency detuning and a complex coupling coefficient are gradually introduced into the model in Sections III-B, III-C and III-D, respectively. In Section IV, we present our results for the case of unequal pumping, demonstrating how increasing $\beta$ yields a much wider range for the steady state values of the phase difference. Finally, Section V concludes the manuscript.

## II. THEORETICAL MODEL AND METHODS

The coupled rate equations, with $\beta$ included, that govern the temporal dynamics of two laterally coupled laser cavities considered here are given by:

$$\frac{d|E_{1,2}|}{d\tau} = \left(\Gamma G_N(N_{1,2} - N_0) - \frac{1}{\tau_p}\right)\frac{|E_{1,2}|}{2}$$
$$+ \frac{\Gamma F_p \beta N_{1,2}}{2\tau_{rad}|E_{1,2}|^2}|E_{1,2}| \quad (1a)$$
$$\mp \kappa \sin(\Delta\Phi)|E_{2,1}| + \gamma \cos(\Delta\Phi)|E_{2,1}|$$

$$\frac{dN_{1,2}}{d\tau} = P_{1,2} - \frac{N_{1,2}}{\tau_{nr}} - \frac{(F_p\beta + 1 - \beta)N_{1,2}}{\tau_{rad}} \quad (1b)$$
$$- G_N(N_{1,2} - N_0)|E_{1,2}|^2$$

$$\frac{d\Delta\Phi}{d\tau} = \frac{\alpha}{2}\Gamma G_N(N_2 - N_1) + \Delta w$$
$$+ \kappa\left(\frac{|E_1|}{|E_2|} - \frac{|E_2|}{|E_1|}\right)\cos(\Delta\Phi) \quad (1c)$$
$$- \gamma\left(\frac{|E_1|}{|E_2|} + \frac{|E_2|}{|E_1|}\right)\sin(\Delta\Phi)$$

where $|E_{1,2}|$ are the amplitudes of the electric fields in cavities 1 and 2, $|E_{1,2}|^2$ are the photon densities, $N_{1,2}$ are the carrier densities and $\Delta\Phi = \Phi_2 - \Phi_1$ is the phase difference between the electric fields in the two cavities. The definitions of the other parameters and their representative values for an InGaAsP material system that is considered in the numerical simulations, are summarized in Table 1. Additionally, even though the Purcell factor $F_p$ and $\beta$ are correlated, they are treated as independent from one another in this study since we are primarily interested in the trends of stability with respect to increasing values of $\beta$.

Finally, the coupling between the two cavities is introduced in a phenomenological manner via a complex coupling coefficient $i\kappa + \gamma$, that includes a dispersive coupling rate $\kappa$ and a dissipative coupling rate $\gamma$. The parameters $\kappa$ and $\gamma$ originate from the dissimilarities in effective refractive indices and losses, respectively, between the two eigenmodes – bonding and anti-bonding - supported by the coupled cavities. To be more precise, $\kappa$ can be calculated from the frequency splitting between the bonding and antibonding modes (denoted by + and -) using $\kappa = \frac{1}{2}(w_+ - w_-)$, while $\gamma$ can be calculated from the loss splitting as $\gamma = \frac{1}{4}\left(\frac{w_-}{Q_-} - \frac{w_+}{Q_+}\right)$, where $Q_{+/-}$ are the quality factors of the supermodes [27]. To generalize the effects of increasing $\beta$ for any laterally coupled system, the dependence of $\kappa$ and $\gamma$ on either coupling geometry or material

## TABLE I
## LASER PARAMETERS

| Symbol | Definition | Value (Unit) |
|---|---|---|
| $\tau_p$ | Photon lifetime | 1.5 $(ps)$ [31,32] |
| $\tau_{rad}$ | Radiative carrier lifetime | 2 $(ns)$ [33] |
| $\tau_{nr}$ | Nonradiative carrier lifetime | 0.625 $(ns)$ [33] |
| $\alpha$ | Henry factor | 4 [32] |
| $\Gamma$ | Confinement factor | 0.8 [31] |
| $N_0$ | Transparent carrier density | $2 \times 10^{24}$ $(m^{-3})$ [32] |
| $F_p$ | Purcell factor | 1 |
| $G_N$ | Differential gain | $4.28 \times 10^{-12}$ $(m^{-3})$ [33, 34] |
| $P_{1,2}$ | Pump rate | $10^{35}$ to $1.2 \times 10^{36} (m^{-3} \cdot s^{-1})$ |
| $\Delta w$ | Frequency detuning between two lasers $\Delta w = w_2 - w_1$ | $-2$ to $2$ $(THz)$ |

properties is neglected, and their values are chosen to be within a range that can be feasibly achieved in coupled laser platforms [26,30]. Although both the sign and the values of $\kappa$ and $\gamma$ can be precisely controlled through altering the coupling geometry and material composition [35-38], such as changing the size of the nanoholes in the center barrier for coupled photonic crystal lasers, we first assume $\gamma = 0$ and $\kappa > 0$ for the simplicity of understanding the model and results presented here. Once we obtain enough initial insight into how stability depends on $\beta$ and the other control parameters, we extend the study to consider a complex coupling coefficient with $\gamma \neq 0$ and the coupling rates demonstrating both positive and negative signs. This allows the model to reflect scenarios where either of the supermodes can exhibit higher eigenfrequency and/or higher losses. In other words, in addition to the coupling geometry and material composition, the sign and values of $\kappa$ and $\gamma$ also depend on the comparative values of the eigenfrequencies and losses of the two supermodes.

## III. PHASE-LOCKING STABILITY V.S. $\beta$

In this study, the stable phase locking regions for the two laterally coupled lasers are identified as functions of $\beta$, pump rate $P$ and the frequency detuning $\Delta w$ using the bifurcation software XPPAUT, which contains the numerical continuation package AUTO [39]. The electric fields and carrier densities in in (1) are normalized to reduce the simulation time. The time scale of the rate equations is also normalized with respect to the photon lifetime $\tau_p$ (see Appendix). Additionally, a small signal analysis is performed to provide physical insight into the results (see Appendix for details). In this work, only three types of bifurcation points are discussed – pitchfork, saddle-node (SN) and Hopf bifurcations – since the stable regions are found to be exclusively bounded by these three types of bifurcations. It is important to note here that although the numerical continuation analysis of the coupled laser model reveals a plethora of dynamical regimes such as stable phase-locking, periodic oscillations, period doubling, and chaotic oscillations, we only consider the conditions that yield stable phase-locking. As a result, the latter three dynamical behaviors are categorized as unstable locking regimes for the purposes of this study.

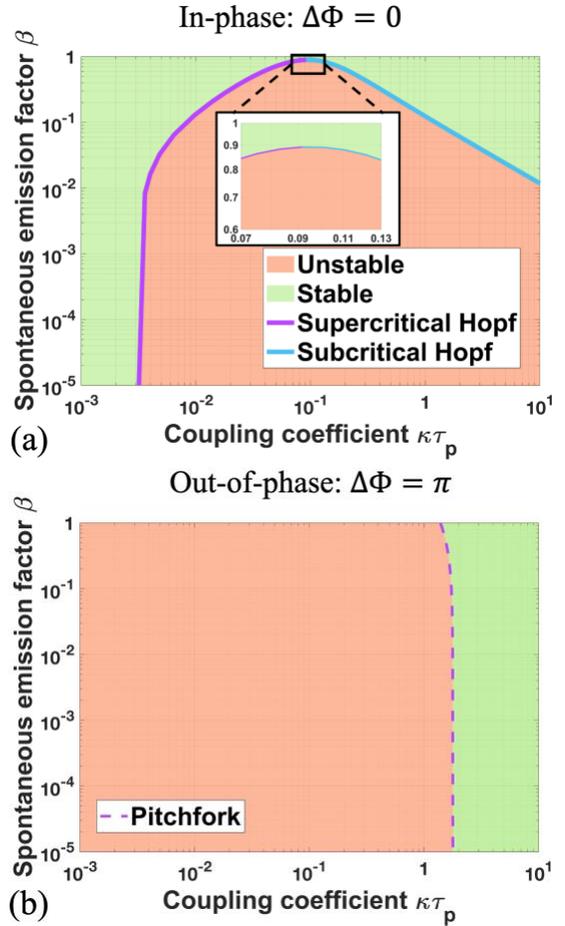

Fig. 1. 2-parameter bifurcation diagrams of the (a) in-phase solution and (b) out-of-phase solution in $(\kappa\tau_p, \beta)$ plane with $P_1 = P_2 = 1.2P_{th}$. Stable locking region is shown in green, unstable region in orange. Solid lines are the supercritical (purple) and subcritical (blue) Hopf bifurcation boundaries. Dashed line is the pitchfork bifurcation boundary.

### A. Imaginary Coupling Coefficient

The simplest representation of the model assumes no frequency detuning, a constant pump rate and a purely imaginary coupling coefficient represented as $i\kappa\tau_p$, which is representative of the case when the two supermodes experience similar losses. Figures 1(a) and (b) illustrate the stability maps within the same parameter space for in-phase ($\Delta\Phi = 0$) and out-of-phase ($\Delta\Phi = \pi$) solutions, respectively, as a function of $\kappa\tau_p$ and $\beta$ at a pump power of $P_{1,2} = 1.2P_{th}$. The variable $P_{th}$ denotes the pump power at lasing threshold for a single laser and can be identified from the steady-state solutions of the rate equations when no coupling is considered. The solid purple and blue lines in Fig. 1(a) denote the Hopf bifurcation boundary and the dashed purple line in Fig. 1(b) represents the pitchfork bifurcation boundary. The Hopf boundary in Fig. 1(a) can be further demarcated into the supercritical Hopf (purple) and the subcritical Hopf (blue) branches. The regions colored in green and orange denote the stable and unstable locking regimes, respectively, for both figures. The coexistence of in-phase and out-of-phase solutions for some values of $\kappa\tau_p$ and $\beta$ can be explained by the fact that the two supermodes exhibit identical losses ($\gamma = 0$).

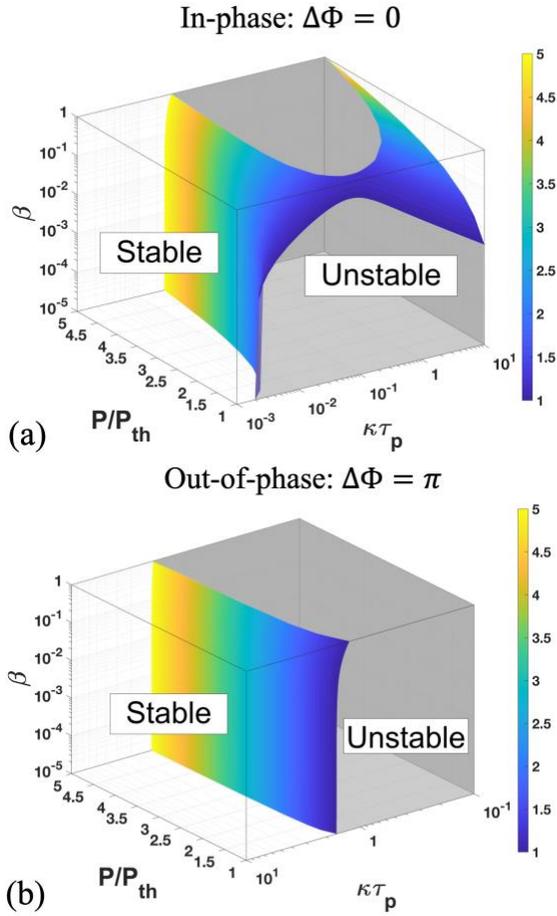

Fig. 2. (a) 3-dimensional stability plot in the $(\kappa\tau_p, P/P_{th}, \beta)$ plane for in-phase solutions. The 3-D surface is the Hopf bifurcation stability boundary. The color denotes various pump rate as shown in the colorbar. Stable phase locking region is shown in white, and the unstable region in grey. (b) 3-D stability plot in the $(\kappa\tau_p, P/P_{th}, \beta)$ plane for out-of-phase solutions using same color convention as in (a). The surface now represents the pitchfork bifurcation stability boundary.

A phenomenon common to both Fig. 1(a) and (b) is the expansion of the green stable regions as $\beta$ is increased from $10^{-5}$ to its maximum possible value of 1. For the out-of-phase solutions depicted in Fig. 1(b), the slight increase in the area of the stable phase-locking region, located to the right of the plot at higher $\kappa\tau_p$ values, is easier to distinguish due to the solitary pitchfork boundary present in this graphic. In comparison, the supercritical and subcritical Hopf bifurcations for in-phase solutions in Fig. 1(a) dissect the parameter space into multiple sections. For $\beta \leq 10^{-2}$, only one stable region exists at very small coupling rates and it remains nearly unchanged in area as $\beta$ increases from $10^{-5}$ to $10^{-2}$, bounded by the supercritical Hopf bifurcation point. In comparison, for $\beta > 10^{-2}$, it can be observed that the stable locking region in the weak coupling regime significantly expands when $\beta$ is increased due to the supercritical Hopf bifurcation point moving to much higher values of $\kappa\tau_p$. Moreover, as the coupling coefficient $\kappa\tau_p$ further increases, a second stable region appears after the subcritical Hopf point, seen towards the right side of Fig. 1(a). This second stability region has not been reported in literature till date, where mainly weak coupling $\kappa\tau_p \ll 1$ and negligible $\beta$ are considered. Despite the complex demarcations in Fig. 1(a), it can be clearly observed that increasing $\beta$ leads to a narrowing of the unstable region as the boundaries of the two Hopf bifurcations move towards each other. In fact, when $\beta = 0.89$, the two bifurcation branches become connected at $\kappa\tau_p \approx 0.09$ as shown in the inset of Fig. 1(a). For values of $\beta$ beyond this point of confluence (i.e. $\beta > 0.89$), the steady-state solutions of the rate equations yield in-phase, stable solutions irrespective of the coupling strength. This result holds major significance as it suggests that nanolasers with $\beta$ values approaching 1 are ideal candidates to be used in phase-locking arrays to generate high power far-field emission.

*B. Pump Rate*

In the previous section, the pump rate was fixed at a constant value for both lasers. In order to gauge whether increasing $\beta$ leads to a similar expansion in the stability regions for any arbitrary pump rate, 3-dimensional (3-D) stability plots with varying $P/P_{th}$ ($P_1 = P_2 = P$) being the third dimension, are created for the in-phase and out-of-phase solutions as shown in Fig. 2(a) and (b), respectively. The stable regions are denoted in white while the unstable ones are shaded in grey in these figures. For the in-phase solutions depicted in Fig. 2(a), when $\beta \leq 0.01$, the supercritical bifurcation branch (surface on the left) moves towards larger $\kappa\tau_p$ as the pump rate $P/P_{th}$ increases, thereby enhancing the stability. This trend has also been reported in another study that focused exclusively on weak coupling and did not consider the effect of the $\beta$ factor [16]. For $\beta \geq 0.01$, however, increasing the pump rate can cause the supercritical bifurcation to shift to smaller $\kappa\tau_p$, and thus shrink the stable locking region. This stability reduction as pump rate increases can be explained by the small signal analysis detailed in the Appendix. Essentially, for weak coupling, the damping rate of the small perturbations can be mathematically approximated to be that of the relaxation oscillations, with this rate increasing for small $\beta$ and decreasing for large $\beta$ as pump rate increases. Therefore, as the pump rate is increased for large $\beta$, the lower damping rate means that the system is now more susceptible to small perturbations and hence, exhibits less stability. Similarly, for the subcritical bifurcation branch (right side of the surface in Fig. 2(a)), an increasing pump rate $P/P_{th}$ pushes the branch to larger coupling coefficients, which also leads to narrowing of the stable locking region. Despite these seemingly disparate effects of the pump on stability, however, the most important observation from Fig. 2(a) is that the two Hopf bifurcation branches always move towards each other as $\beta$ increases. The increasing proximity of the two bifurcations in turn, results in an expansion of the stability region. Therefore, it can be concluded that although varying the pump rate affects the stability in a non-uniform manner, higher $\beta$ values always contribute towards increased in-phase locking irrespective of the pump rate.

In contrast, for the out-of-phase solutions depicted in Fig. 2(b), the pitchfork bifurcation boundary remains almost unaltered despite varying both $P/P_{th}$ and $\beta$. The reason for this can be inferred from small signal analysis (see Appendix), which reveals that the pitchfork boundary is approximately proportional to $N_{1,2} - N_0$. Given that $N_{1,2}$ clamps to a threshold value as $P/P_{th}$ increases, the stability boundary therefore stays almost constant. Though higher $\beta$ values result in a slight increase of $N_{1,2}$, the consequent expansion of the stability region is miniscule. Fortuitously, for most applications, only the in-phase solutions are of interest as they are essential for the generation of higher power density. Therefore, in the next section when we consider frequency detuning, we focus exclusively on the solutions around $\Delta\Phi = 0$, which are referred to as "in-phase solutions" for simplicity.

## C. Frequency Detuning

Now we consider the case where two cavities exhibit disparate resonance frequencies and investigate whether a stability enhancement from high-$\beta$ can be observed in this scenario. While frequency detuning is usually induced by dissimilarities in the dimensions of the resonators caused by fabrication imperfections, it can also be intentionally introduced into the coupled structure for certain applications. For example, phase-locked laser arrays with shifted frequencies between the adjacent elements can be implemented in ultra-high-resolution lidar systems for distance-angle beam steering tasks [40-42].

In Fig. 3, the stable regions of the in-phase solutions are depicted in a 3-D parameter space $(\kappa\tau_p, \Delta w\tau_p, \beta)$ with $P_{1,2}/P_{th} = 1.2$. To provide a more intuitive visualization, the parameter space is dissected into two regions at $\kappa\tau_p = 0.1$, which is the approximate point of confluence where the supercritical and subcritical Hopf branches become connected, as shown in the zoomed-in inset of Fig. 1(a). Consequentially, Fig. 3(a) represents the region of $\kappa\tau_p \leq 0.1$ containing the supercritical Hopf bifurcation while Fig. 3(b) illustrates the scenario when $\kappa\tau_p \geq 0.1$ and the subcritical Hopf branch is observed. Like in Fig. 2, the stable and unstable regions are represented in white and grey, respectively, in Fig. 3(a) and (b) as well.

Detuning the frequency gives rise to two symmetric SN bifurcation boundaries for the case of weak coupling ($\kappa\tau_p \leq 0.1$) depicted in Fig. 3(a). These SN bifurcation surfaces, along with the supercritical Hopf boundary, enclose the stable in-phase locking region. As detuning is decreased, the SN boundaries move closer to one another but remain unconnected for the case of zero detuning resulting in only Hopf bifurcation boundaries that are observed in this case. More importantly, as $\beta$ increases, although the SN boundaries remain largely unperturbed, the supercritical Hopf branch relocates to higher $\kappa\tau_p$ values. This, in turn, expands the stable phase locking region in Fig. 3(a). Similarly, for $\kappa\tau_p \geq 0.1$ in Fig. 3(b), the stable phase locking region is also seen to expand for higher $\beta$ values when the subcritical Hopf bifurcation serving as its sole boundary shifts towards smaller $\kappa\tau_p$. It is important to note here that the subcritical Hopf points for ultra-small $\beta$ (the bluer-parts of the 3-D surface in Fig. 3(b)) require extremely high values of $\kappa\tau_p$ ($\kappa\tau_p > 10$) which are not realistically achievable in experiment. This explains why the second stability region shown in Fig. 3(b) has not been previously reported in the literature where usually, only weak coupling and negligible $\beta$ are considered. The results here indicate that increasing $\beta$ helps expand the stable, in-phase locking regions for both weak and strong coupling cases despite the lasers demonstrating dissimilar frequencies.

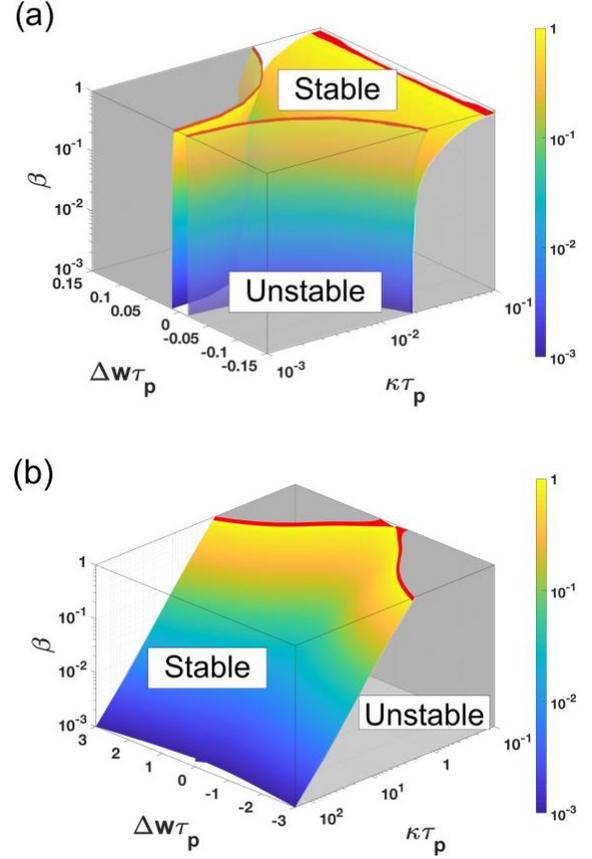

Fig. 3. (a) 3-dimensional stability plot in the $(\kappa\tau_p, \Delta w\tau_p, \beta)$ plane for in-phase solutions with $\kappa\tau_p \in [10^{-3}, 10^{-1}]$. The 3-D surfaces are the stability boundaries for Hopf and SN bifurcations. The colors denote varying $\beta$ as shown in the colorbar. The red region denotes $\beta \geq 0.89$. Stable phase locking region is shown in white, and the unstable region is colored in grey. (b) 3-D stability plot in the $(\kappa\tau_p, \Delta w\tau_p, \beta)$ plane for in-phase solutions with $\kappa\tau_p \in [10^{-1}, 200]$ using the identical color convention as in (a).

In Fig. 1(a), it was seen that when $\beta \geq 0.89$, stability holds irrespective of the strength of coupling for the $\Delta w = 0$ case. However, the same does not hold true when the frequency detuning between the two lasers is non-zero. For $\beta \geq 0.89$ with non-zero detuning, the stability boundaries are colored in red as shown in Fig. 3(a) and (b) and indicate that the stability is lost when detuning is non-negligible. To better illustrate the in-phase stability map with frequency detuning in the low, moderate and high $\beta$ regimes, we combine the parameter spaces shown in Fig. 3(a) and (b) and present them in Fig. 4 as 2-D parameter projections at $\beta = 10^{-3}, 0.25$ and approaching 1. The stable regions are now denoted in green while their unstable counterparts are colored in orange. Considering first the case of $\beta = 10^{-3}$ in Fig. 4(a), the stable solutions can be seen to be enclosed in an extremely narrow parameter space by the SN (dashed purple lines) and supercritical Hopf (solid purple line) bifurcations. As $\beta$ is increased to 0.25 in Fig. 4(b), the stable region expands to cover a much larger area while a second stable region is created at high $\kappa\tau_p$ values due to the presence of the subcritical Hopf boundary (solid blue line). Finally, as $\beta$ is increased beyond 0.89 in Fig. 4(c), the supercritical and subcritical Hopf boundaries merge, resulting in stable in-phase locking regions that span a significantly larger range of both $\kappa\tau_p$ and $\Delta w\tau_p$ values. Akin to what was observed when varying the pump rate, the desirable result of high-$\beta$ increasing the stability of two laterally coupled lasers is preserved even when frequency detuning is considered.

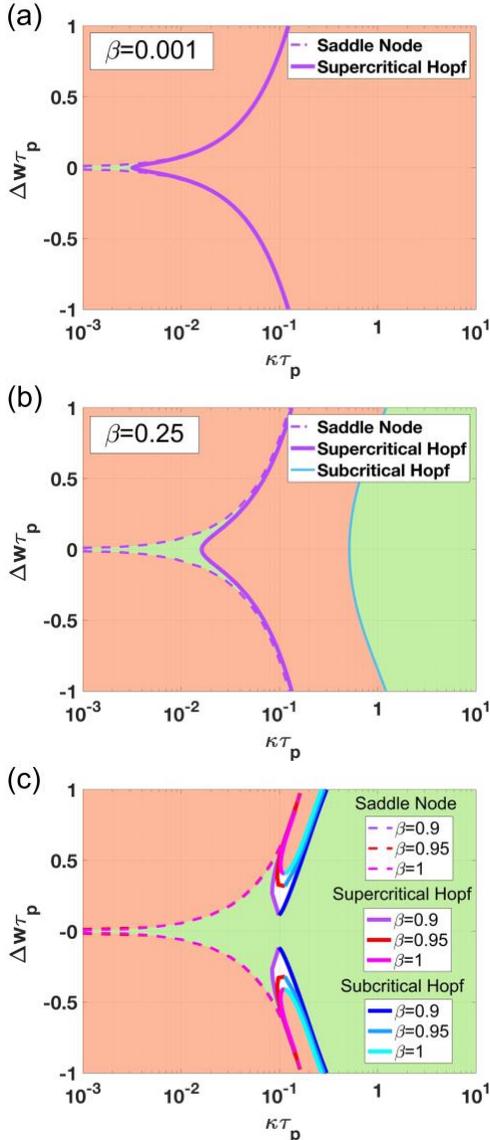

Fig. 4. 2-parameter bifurcation diagrams of the in-phase solutions in the $(\kappa\tau_p, \Delta w\tau_p)$ plane with $P_1 = P_2 = 1.2 P_{th}$ for (a) $\beta = 10^{-3}$, (b) $\beta = 0.25$, and (c) $\beta \geq 0.9$. Stable locking regions are colored in green, unstable regions in orange. Solid lines represent the supercritical (purple, red, magenta) and subcritical (blue, light blue, cyan) Hopf bifurcation boundaries. Dashed lines denote the SN bifurcation points.

It is worth mentioning here that the enhancement of stability due to large $\beta$ for the non-zero frequency detuning case is not restricted only to the pump rate assumed in the above results. Though they have not been included in this work, additional simulations show that increasing or decreasing the pump rate around $P_{1,2}/P_{th} = 1.2$ will only slightly modify the quantitative value of the bifurcation points while the main features of the stability plots remain unaltered. Specifically, increasing the pump rate provides increased stability for coupled lasers with small $\beta$ and a reduction in the stable region for coupled lasers with high $\beta$, as observed in Fig. 2(a). More importantly, it is observed that for any given pump rate, systems with higher $\beta$ always demonstrate a larger stable phase locking region over the parameter space $(\kappa\tau_p, \Delta w\tau_p)$, i.e. better stability. Another reason for choosing the pump rate of $P_{1,2}/P_{th} = 1.2$ for the above simulations is that in practice, lasers operating lightly above threshold have the highest energy transfer efficiency, i.e. wall plug efficiency (WPE) [43], and can also be prevented from overheating due to large current injection. Since nanolasers with high $\beta$ exhibit a low pumping threshold [44], it is not only energy efficient to operate slightly above threshold, but stable phase locking is also most achievable with a high $\beta$ value and low pump rate.

### D. Complex coupling coefficient

To further extend the analysis to account for the scenarios where the supermodes exhibit dissimilar losses, a complex coupling coefficient $i\kappa + \gamma$ - where $\kappa$ and $\gamma$ can be either positive or negative - is used to replace the purely imaginary coupling coefficient used thus far in the model. This modification is especially important if considering coupling geometries employing gain-guiding and carrier-induced index antiguiding [16], where either of the supermodes can exhibit higher eigenfrequencies and/or higher losses. In order to simplify the ensuing bifurcation analysis, the coupling amplitude $|\kappa|$ and phase $\theta_\kappa$ are used instead of $\kappa$ and $\gamma$ to provide better intuition for the control parameters used, i.e. $i|\kappa|e^{j\theta_\kappa} = i\kappa + \gamma, \kappa = |\kappa|\cos\theta_\kappa$, and $\gamma = -|\kappa|\sin\theta_\kappa$, where $\theta_\kappa \in [-\pi, \pi]$. Furthermore, the pump rates are assumed to be $P_{1,2}/P_{th} = 1.2$ to obtain a high energy efficiency in practice. Altering the pump rate around this value does not significantly impact the general shape of the stability regions and only negligibly shifts the boundaries. Therefore, the variation of the stable phase locking regions due to $\beta$ is unlikely to be affected by the choice of the pump rate level. Finally, the frequency detuning is assumed to be 0 initially for simplicity, with a more detailed analysis on non-zero detuning discussed briefly towards the end of this section.

The stability plots, when considering the complex coupling coefficient and varying $\beta = 10^{-3}$, 0.05, 0.25 and 1, are superimposed and presented in Fig. 5(a). Although the bifurcation analysis yields a plethora of bifurcation boundaries, in Fig. 5(a), we only show those that directly demarcate the stable and unstable regions, i.e. Hopf points (solid lines) and the pitchfork points (dashed lines). In Fig. 5(a), the regions in red represent stable in-phase solutions, the ones colored in blue denote stable out-of-phase solutions and the white, unshaded regions represent unstable solutions. From this figure, it can be observed that the stability regions of in-phase (red) and out-of-phase solutions (blue) with the same $\beta$ values are identical in shape albeit shifted with respect to each other by $\pi$ radians. The reason for this horizontal shift, which can be easily deduced from the rate equations (Appendix (A1)), is that if $\gamma$ flips its sign, i.e. $\theta_\kappa$ becomes $\theta_\kappa + \pi$, then $\Delta\Phi$ is shifted by $\pi$ radians. This underlines the significance that the signs of $\kappa$ and $\gamma$ hold and how controlling them would allow a coupled system to achieve the desired phase difference as predicted by the rate equations. However, the even more significant finding from Fig. 5(a) is that as $\beta$ increases, both the in-phase and out-of-phase solutions expand in size, which is consistent with what was observed in Fig. 1(a) and (b). We would like to note that not all coupling values illustrated in Fig. 5(a) are realistically achievable in experiment. For instance, for coupled systems with $Q_{+/-}$ on the order of hundreds of more, $\gamma$ cannot be on the same order as $\kappa$ and therefore $\theta_\kappa$ values around $\pm\pi/2$ cannot be realized from the definition of the coupling coefficient provided in Section II. Nevertheless, the purpose of choosing this range of complex coupling coefficients is to provide an accurate picture of how the stability regions expand as $\beta$ is increased.

Another interesting observation is the coexistence of in-phase and out-of-phase solutions in certain parts of the $(\theta_\kappa, |\kappa\tau_p|)$ parameter space, specifically around $\theta_\kappa = 0$ (and $\theta_\kappa = \pi$). Recall that around these values, the supermodes have nearly identical losses and therefore exhibit approximately equal probability of being supported by the coupled system. The same coexistence of solutions was also observed in Fig. 1(a) and (b), since the purely imaginary and positive coupling coefficient used in the analysis in that section can be viewed as a special case of $\theta_\kappa = 0$. The evolution of these bistable regions as $\beta$ is varied is plotted in Fig. 5(b)-(e). It can be clearly observed that in addition to expanding stable regions, increasing $\beta$ can also lead to a larger overlap of the in-phase and out-of-phase stable solutions, thereby increasing the likelihood of achieving bistability. Within these bistable regions, the final steady state depends on the initial state of the phase relations between the two solutions. Such bistable operation poses great potential for use as memories such as for optical flip-flops [27, 45] as well as for optical analogues of the degenerated spins in an Ising machine [46].

When non-zero frequency detuning is considered along with the complex coupling coefficient, it is observed that both the in-phase and out-of-phase stable regions have a lower bound in $|\kappa\tau_p|$, which is due to the SN boundary arising from the non-zero $\Delta w$. As $\Delta w$ increases, both the SN and Hopf bifurcation points are shifted in a manner that reduces the stable phase locking region in the parameter space $(\theta_\kappa, |\kappa\tau_p|)$. However, for any non-zero $\Delta w$, an enhanced stability from higher $\beta$ can always be observed, which is consistent with the results from previous sections. Though discussed briefly here, the detailed results are not included in this work for brevity.

In summary, with regards to stable phase-locking, increasing $\beta$ unequivocally leads to an expansion in the stability regions despite considering varying pump rates, detuned frequencies and both imaginary and complex coupling coefficients. The robustness of the desirable effects of high $\beta$ on stability truly emphasize the tantalizing potential of nanolaser arrays to harness this advantage and help in the generation of high optical power via in-phase locking. Additionally, high-$\beta$ nanolaser arrays can also aid in the development of next generation active optical phased arrays as discussed in the next section.

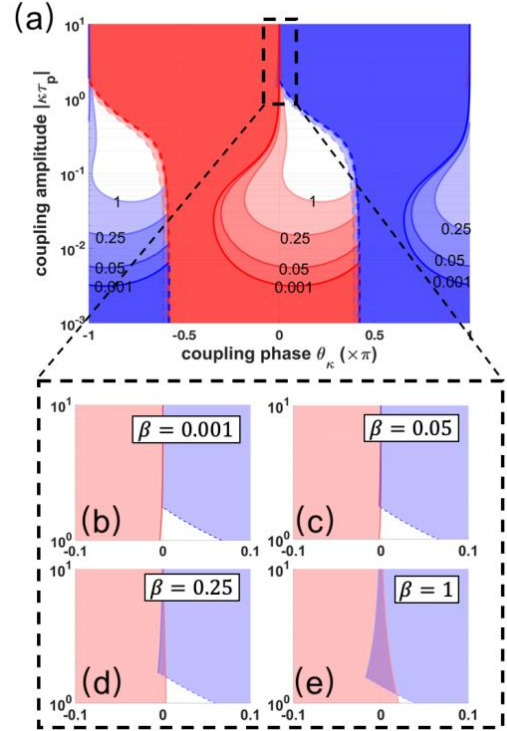

Fig. 5. (a) 2-parameter bifurcation diagrams of in-phase and out-of-phase solutions in the $(\theta_\kappa, |\kappa\tau_p|)$ plane with $\beta = 10^{-3}, 0.05, 0.25$ and $1$. The stable in-phase locking region is shown in red; the stable out-of-phase locking region is colored in blue and the unstable region is shown in white. The solid lines denote Hopf bifurcations, while the dashed lines denote Pitchfork bifurcations. (b) – (e) Zoom-in of the region $\theta_\kappa \in [-0.1\pi, 0.1\pi], |\kappa\tau_p| \in [1,10]$ for $\beta = 0.001, 0.05, 0.25, 1$, respectively.

## IV. Phase Difference Modulation V.S. $\beta$

In the previous section, the stability of phase locking was studied as functions of $\beta$, $P$, $\Delta w$ and $\theta_\kappa$. The results were focused on the in-phase ($\Delta\Phi = 0$) and out-of-phase ($\Delta\Phi = \pi$) solutions for their potential in high-power beam generation and optical memory. In some other applications such as beam steering for Lidar and imaging systems, a tunable phase offset between adjacent lasers is required. In fact, having a wide range of tunable phase differences between coupled lasers can prove essential in these applications, since this attribute can help increase the azimuthal and vertical scanning ranges. Using lasers as array elements instead of passive phase shifters injected by a single laser source offers the advantage of both frequency and phase reconfiguration, which are essential for complex detection and sensing applications [40, 41]. In this section, we theoretically propose and analyze a method to modulate the phase difference between two coupled lasers. For a symmetrically coupled system like we have considered thus far, i.e. equal pumping rate $P_1 = P_2$, the case of zero-frequency detuning $\Delta w = 0$ yields only two possible solutions for the steady-state phase difference $\Delta\Phi$: the in-phase ($\Delta\Phi = 0$) and the out-of-phase solutions ($\Delta\Phi = \pi$). If the symmetry between the two lasers is broken by pumping the cavities at dissimilar rates, then values of $\Delta\Phi$ that are neither 0 nor $\pi$ are achievable. In fact, $\Delta\Phi$ can then be tuned within the stable phase-locking range according to the ratio of the pump rates for the two lasers.

To identify the feasibility of nanolasers to be implemented in novel phased arrays for beam steering, the dependence of the phase difference tunability on $\beta$ is investigated. In the simulation for each $\beta$ value, the pump rate for one of the lasers, $P_1$, is fixed while the pump rate for its neighbor, $P_2$, is varied. We choose to keep $P_1/P_{th} = 1.2$ for the same reason of energy efficiency that was mentioned in the previous sections. To realize phase difference modulation, $P_2$ needs to be varied within a range where only stable phase locking is supported by the coupled cavities. Additionally, $P_2$ needs to be experimentally achievable and is thus varied only from $P_{th}$ to $12P_{th}$ throughout this simulation. The three sequential steps followed to perform the analysis are as follows: First, by keeping $\beta$ and $P_1$ constant, a one-parameter bifurcation analysis by varying $P_2$ is conducted, and the maximum and minimum possible $\Delta\Phi$ within the stable region are recorded. Secondly, the above step is repeated for $\beta$'s ranging from $10^{-5}$ to 1, and the maximum and minimum $\Delta\Phi$ that can be achieved by varying $P_2$ are recorded for each $\beta$ value. Finally, these results are depicted in Fig. 6(a) where the maximum and minimum $\Delta\Phi$ values are plotted as a function of $\beta$, as well as in Fig. 6(b) where the phase tuning range representing the differences between the maximum and minimum $\Delta\Phi$ is also shown as a function of $\beta$. For these simulations, the frequency detuning between the two lasers is neglected, and $\gamma$ is assumed to be 0 for simplicity. Additionally, only an imaginary and constant coupling rate of $\kappa\tau_p = 10^{-3}$ is considered. It is important to note here that phase tunability was only observed with a coupling rate within the first stability region (to the left of the supercritical Hopf bifurcation boundary in Fig. 1(a)) and not for $\kappa\tau_p$ values in the second stability region (to the right of the subcritical Hopf boundary in Fig. 1(a)). Moreover, within the first stability region, varying $\kappa\tau_p$ affects the values of $\Delta\Phi$ only in a negligible manner.

As can be clearly observed in Fig. 6(a) and (b), as $\beta$ is increased, a wider range of tuning in $\Delta\Phi$ is afforded. Specifically, in the yellow region demarcated by extremely low-$\beta$ ($10^{-5}$ to $10^{-3}$), the maximum and minimum $\Delta\Phi$ achievable are around $-0.05\pi$ and $0.05\pi$, respectively. As $\beta$ is increased to values in the blue region, the range for $\Delta\Phi$ expands significantly to about $[-\pi/2, \pi/2]$. Therefore, the phase tuning range shown in Fig. 6(b) increases from around $0.1\pi$ to $\pi$ as $\beta$ increases from that of conventional semiconductor lasers, i.e. $\beta \leq 10^{-3}$, to that of microscale and nanoscale lasers, i.e. $\beta > 0.01$. The reason for this wider range of phase tunability brought about by increasing $\beta$ lies in the manner in which bifurcation points alter the stable solutions. For the range of extremely low-$\beta$ values shaded as the yellow region in Fig. 6(a), the coupled lasers remain stable for all values of $P_2/P_{th} \in [1,12]$ as shown in Fig. 6(c). However, when $\beta$ is increased to values in the pink region of Fig. 6(a), an SN bifurcation point arises that pushes the lower limit of $\Delta\Phi$ closer to $-\pi/2$. This result is encapsulated in Fig. 6(d) for a specific value of $\beta =$ 0.0175 that lies within the pink region in Fig. 6(a). Finally, for high-$\beta$ values in the blue region of Fig. 6(a), two SN bifurcation points further define the stability boundary such that $\Delta\Phi$ can now vary from $-\pi/2$ to $\pi/2$ when $P_2/P_{th}$ varies within a small range around 1.2, as illustrated in Fig. 6(e) for $\beta = 0.1$. Therefore, increasing $\beta$ can significantly increase the range of phase differences possible for stable phase-locked solutions, highlighting the fact that laterally coupled nanolasers with intrinsically high $\beta$ values can prove valuable in realizing wide scanning angles in optical phased arrays.

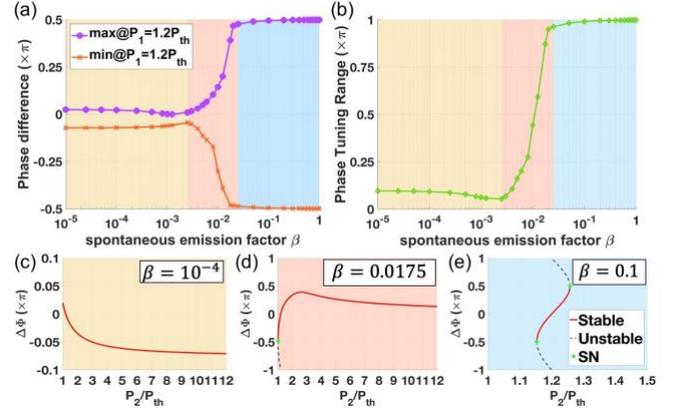

Fig. 6. (a) Maximum (purple) and minimum (orange) phase differences achieved by varying the pump rate $P_2$ while keeping $P_1$ constant, plotted as a function of $\beta$. For each $\beta$ value, the maximum values are marked with circles, and the minimum values are marked with asterisks. $P_1$ is set to be $1.2P_{th}$. (b) Phase difference tuning range as a function of $\beta$. The colored regions in both (a) and (b) represent the number of bifurcation points observed in the solutions. This is better illustrated in (c)-(e) which show the steady-state phase differences for three different values of $\beta$- $10^{-4}$, 0.0175, and 0.1. Depending on the value of $\beta$, there may exist zero (c), one (d) or two (e) SN bifurcations points in the solution, corresponding to the yellow, pink and blue regions in (a), respectively.

## V. Conclusion

The theoretical effects of varying the spontaneous emission factor, $\beta$, on the stability and tunability of phase-locking in two laterally coupled semiconductor lasers are presented in this study. In order to first determine how $\beta$ affects the stability of the coupled system, bifurcation analysis is performed over the laser rate equations using numerical continuation. Initial results with a simplistic model considering constant and equal pump rates, identical resonance frequencies and an imaginary coupling coefficient reveal that increasing $\beta$ leads to an overall expansion of the stable phase-locking regions. To account for realistic experimental conditions and practical device designs, additional control parameters such as varying pump rate, frequency detuning and complex coupling coefficients were considered in the model. The desirable effects of high-$\beta$ on stability were found to be robust to the addition of these multiple parameters. More importantly, the stable in-phase locking regions, conducive for generating high output optical power, were observed to increase in area as a direct result of increasing $\beta$. Such stability enhancement becomes even more significant for $\beta \geq 0.89$, where the in-phase solutions are stable over a wide range of coupling coefficients and frequency detuning. During the stability analysis, regions of bistability that increase in area due to increasing $\beta$ were also observed. The simultaneous coexistence of two solutions in this manner can find applications in optical memories. Finally, higher values of $\beta$ were also found to exert influence on the range of stable phase differences attainable from a laterally coupled system. By breaking the symmetry of pumping for the two lasers while altering $\beta$, a range of phase differences as wide as $\pi$ (from $-\pi/2$ to $\pi/2$) was attained for $\beta > 0.025$. Wide ranges of tunability of this form are desirable in applications that require a large scanning angle and beam steering such as in lidar systems. This is the first study demonstrating rigorous analysis on the specific effects of high-$\beta$ on the phase-locking stability and tunability of laterally coupled semiconductor lasers, to the best of our knowledge. Future analysis on the stability of coupled lasers can consider more than just two interacting lasers as well as specific coupling geometries.

## APPENDIX

### A. Normalization and the Linear Analysis of Coupled Rate Equations

Before we perform the small signal analysis, the different bifurcation points need to be introduced and identified. The saddle-node bifurcation indicates the collision and disappearance of two equilibria. A pitchfork bifurcation occurs when the system transitions from one fixed point to three fixed points. In both these types of bifurcation points, the Jacobian matrix of the dynamical systems has one zero eigenvalue. In contrast, at the Hopf bifurcation points, the solution switches from being stable to exhibiting periodicity, i.e. instability. For the supercritical Hopf bifurcation, one fixed point diverges into periodic oscillations, while the reverse holds true for the subcritical Hopf bifurcation. The occurrence of this Hopf point corresponds to a pair of purely imaginary eigenvalues.

We use a linear gain model for $G(N_{1,2})$ with $G(N_{1,2}) = G_N(N_{1,2} - N_0)$, where $G_N$ is the differential gain and $N_0$ is the carrier density at transparency. The rate equations in (1) can be normalized using $X_{1,2} = |E_{1,2}|\sqrt{G_N \tau_{nr}}$, $Y_{1,2} = (N_{1,2} - N_0)\Gamma \tau_p G_N$ and a dimensionless time that is normalized to the photon lifetime as $t = \tau/\tau_p$. We can then write the normalized equation as:

$$\frac{dX_{1,2}}{dt} = \frac{1}{2}(Y_{1,2} - 1)X_{1,2} + \frac{\tau_{nr}}{\tau_{rad}} \frac{F_p \beta}{2} \frac{(Y_{1,2} + N_{0_{norm}})}{|X_{1,2}|^2}|X_{1,2}|$$
$$\mp \kappa \tau_p \sin(\Delta\Phi)X_{2,1} + \gamma \tau_p \cos(\Delta\Phi) X_{2,1} \quad (A1a)$$

$$\frac{dY_{1,2}}{dt} = T_{norm}[P_{norm} - \gamma_c(Y_{1,2} + N_{0_{norm}}) - Y_{1,2}X_{1,2}^2] \quad (A1b)$$

$$\frac{d\Delta\Phi}{dt} = \frac{\alpha}{2}(Y_2 - Y_1) + \Delta w \tau_p + \kappa \tau_p \left(\frac{X_1}{X_2} - \frac{X_2}{X_1}\right)\cos(\Delta\Phi)$$
$$- \gamma \tau_p \left(\frac{X_1}{X_2} + \frac{X_2}{X_1}\right)\sin(\Delta\Phi) \quad (A1c)$$

where $T_{norm} = \tau_p/\tau_{nr}$, $N_{0_{norm}} = N_0 \Gamma \tau_p G_N$, $\gamma_c = \frac{\tau_{nr}}{\tau_{rad}}(F_p \beta + 1 - \beta) + 1$, and $P_{norm} = P_{norm}\Gamma \tau_p G_N/\tau_{nr}$ is the normalized pump rate.

We then perform small signal analysis like in ref. 16 and assume:
$$X_{1,2} = \bar{X}_{1,2} + x_{1,2}e^{\lambda t}, Y_{1,2} = \bar{Y}_{1,2} + y_{1,2}e^{\lambda t},$$
$$\Delta\Phi = \Delta\bar{\Phi} + \delta\phi e^{\lambda t} \quad (A2)$$

Substituting $(A2)$ into $(A1)$, neglecting higher order terms and assuming $\overline{X_1} \approx \overline{X_2}$, results in:

$$x_1 \lambda = \frac{1}{2}(\bar{Y}_1 - 1)x_1 - \frac{\tau_{nr}}{\tau_{rad}} \frac{F_p \beta}{2} \frac{(Y_1 + N_{0_{norm}})}{|\bar{X}_1|^2}x_1$$
$$+ \left(\frac{1}{2}\bar{X}_1 + \frac{\tau_{nr}}{\tau_{rad}} \frac{F_p \beta}{2} \frac{1}{\bar{X}_1}\right)y_1$$
$$+ [\gamma \tau_p \cos(\Delta\bar{\Phi}) - \kappa \tau_p \sin(\Delta\bar{\Phi})]x_2$$
$$- [\gamma \tau_p \sin(\Delta\bar{\Phi}) + \kappa \tau_p \cos(\Delta\bar{\Phi})]\bar{X}_2 \delta\phi \quad (A3a)$$

$$x_2 \lambda = \frac{1}{2}(\bar{Y}_2 - 1)x_2 - \frac{\tau_{nr}}{\tau_{rad}} \frac{F_p \beta}{2} \frac{(Y_2 + N_{0_{norm}})}{|\bar{X}_2|^2}x_2$$
$$+ \left(\frac{1}{2}\bar{X}_2 + \frac{\tau_{nr}}{\tau_{rad}} \frac{F_p \beta}{2} \frac{1}{\bar{X}_2}\right)$$
$$+ [\gamma \tau_p \cos(\Delta\bar{\Phi})_- + \tau_p \sin(\Delta\bar{\Phi})]x_1$$
$$- [\gamma \tau_p \sin(\Delta\bar{\Phi}) - \kappa \tau_p \cos(\Delta\bar{\Phi})]\bar{X}_1 \delta\phi \quad (A3b)$$

$$y_1 \lambda = T_{norm}\left(-\gamma_c y_1 - 2\bar{X}_1 \bar{Y}_1 x_1 - \bar{X}_1^2 y_1\right) \quad (A3c)$$

$$y_2 \lambda = T_{norm}\left(-\gamma_c y_2 - 2\bar{X}_2 \bar{Y}_2 x_2 - \bar{X}_2^2 y_2\right) \quad (A3d)$$

$$\delta\phi \lambda = \frac{\alpha}{2}(y_2 - y_1) + 2\kappa \tau_p \cos(\Delta\bar{\Phi})\frac{x_1 - x_2}{\bar{X}_1}$$
$$- 2\gamma \tau_p \sin(\Delta\bar{\Phi})\frac{x_1 + x_2}{\bar{X}_1} - 2\gamma \tau_p \cos(\Delta\bar{\Phi})\delta\phi \quad (A3e)$$

By considering equal pumping and neglecting the dissipative coupling by setting $\gamma = 0$, the above equations can be further simplified. We then add $(A3a)$ and $(A3b)$, as well as $(A3c)$ and $(A3d)$ and arrive at:

$$(x_1 + x_2)\left[\lambda - \frac{1}{2}(\bar{Y}_1 - 1) + \frac{\tau_{nr}}{\tau_{rad}} \frac{F_p \beta}{2} \frac{(Y_1 + N_{0_{norm}})}{|\bar{X}_1|^2}\right]$$
$$= \left(\frac{1}{2}\bar{X}_1 + \frac{\tau_{nr}}{\tau_{rad}} \frac{F_p \beta}{2} \frac{1}{\bar{X}_1}\right)(y_1 + y_2) \quad (A4a)$$

$$(x_1 + x_2)(-2T_{norm}\bar{X}_1 \bar{Y}_1)$$
$$= [\lambda + T_{norm}(\gamma_c + \bar{X}_1^2)](y_1 + y_2) \quad (A4b)$$

Combining $(A4a)$ and $(A4b)$, we have:
$$\lambda^2 + A_1 \lambda + A_2 = 0$$

where

$$A_1 = T_{norm}(\gamma_c + \bar{X}_1^2) - \frac{1}{2}(\bar{Y}_1 - 1) + \frac{\tau_{nr}}{\tau_{rad}} \frac{F_p \beta}{2} \frac{(Y_1 + N_{0_{norm}})}{|\bar{X}_1|^2}$$

Recall that $\frac{\tau_{nr}}{\tau_{rad}} \frac{F_p \beta}{2} \frac{(Y_1 + N_{0_{norm}})}{|\bar{X}_1|^2} = -\frac{1}{2}(\bar{Y}_1 - 1)$ and $A_1 = T_{norm}(\gamma_c + \bar{X}_1^2) - (\bar{Y}_1 - 1)$. In order for the small perturbations to approach zero as time evolves, the real part of $\lambda$ must be negative. This requires $A_1 = 2Re(\lambda) > 0$. For the expression of $A_1$, the first term on the RHS signifies the radiative recombination of carriers by all means, and the second term denotes recombination involving only spontaneous emission. Therefore, $A_1 > 0$ always hold true.

We then subtract $(A3b)$ from $(A3a)$, as well as $(A3d)$ from $(A3c)$ and arrive at:

$$(x_1 - x_2)\left[\lambda - \frac{1}{2}(\bar{Y}_1 - 1) + \frac{\tau_{nr}}{\tau_{rad}}\frac{F_p\beta}{2}\frac{(Y_1 + N_{o\,norm})}{|\bar{X}_1|^2}\right]$$
$$= \left(\frac{1}{2}\bar{X}_1 + \frac{\tau_{nr}}{\tau_{rad}}\frac{F_p\beta}{2}\frac{1}{\bar{X}_1}\right)(y_1 - y_2)$$
$$-2\kappa\tau_p \cos(\Delta\bar{\Phi})\bar{X}_1\delta\phi \quad (A5a)$$

$$(x_1 - x_2)(-2T_{norm}\bar{X}_1\bar{Y}_1)$$
$$= [\lambda + T_{norm}(\gamma_c + \bar{X}_1^2)](y_1 - y_2) \quad (A5b)$$

Substituting $(A3e)$ into $(A5a)$ results in:
$$\lambda^3 + B_1\lambda^2 + B_2\lambda + B_3 = 0 \quad (A6)$$

where
$$B_1 = T_{norm}(\gamma_c + \bar{X}_1^2) - (\bar{Y}_1 - 1) \quad (A7a)$$
$$B_2 = T_{norm}\bar{X}_1\bar{Y}_1\left(\bar{X}_1 + \frac{\tau_{nr}}{\tau_{rad}}F_p\beta\frac{1}{\bar{X}_1}\right)$$
$$-T_{norm}(\gamma_c + \bar{X}_1^2)(\bar{Y}_1 - 1) + 4\kappa^2\tau_p^2\cos^2(\Delta\bar{\Phi}) \quad (A7b)$$
$$B_3 = 4\kappa^2\tau_p^2\cos^2(\Delta\Phi)T_{norm}(\gamma_c + \bar{X}_1^2)$$
$$+2\alpha\kappa\tau_p T_{norm}\bar{X}_1^2\bar{Y}_1\cos(\Delta\bar{\Phi}) \quad (A7c)$$

The solutions to $(A6)$ are one real value and two conjugate complex values. The real solution gives the saddle-node bifurcations or pitchfork bifurcations while the complex solutions gives the Hopf bifurcations.

For very weak coupling, $B_3$ in $(A7c)$ is approximately 0 and $(A6)$ can be simplified to be quadratic. Consequentially, the solution of $\lambda$ can then be approximated to be that of the relaxation oscillations (RO), where $Re(\lambda)$ is the damping rate and $Im(\lambda)$ is the RO frequency, which is $T_{norm}(\gamma_c + \bar{X}_1^2) + \frac{\tau_{nr}}{\tau_{rad}}F_p\beta\frac{(Y_1+N_{o\,norm})}{|\bar{X}_1|^2}$. In this case, a larger $\beta$ always results in faster damping, therefore, enhancing the stability in the weak coupling region. The pump rate can also increase the damping rate for small $\beta$. For large $\beta$, the scenario becomes more complex and requires more detailed examination. However, since the damping rate can be approximated to be that of the RO as $T_{norm}(\gamma_c + \bar{X}_1^2) + \frac{\tau_{nr}}{\tau_{rad}}F_p\beta\frac{(Y_1+N_{o\,norm})}{|\bar{X}_1|^2}$, for very small $\beta$, the second term can be neglected. This means that as pump rate increases, a larger $\bar{X}_1^2$ gives a faster damping, i.e. better stability. While for larger $\beta$, the second term can not be neglected. Since $\bar{X}_1^2$ is now in the denominator, a larger $\bar{X}_1^2$ gives a slower damping rate, i.e. a worse stability.

To have the real parts of the solutions to $\lambda$ be negative, and thus have stable phase locking, the following conditions must hold,
$$B_1 > 0, B_3 > 0, B_1B_2 - B_3 > 0 \quad (A8)$$

Since $B_1 = A_1 > 0$ has already been proven to be true, we focus on the second and the third conditions.

The condition $B_3 > 0$ makes the real solution negative, and thus yields,
$$4\kappa^2\tau_p^2\cos^2(\Delta\bar{\Phi})T_{norm}(\gamma_c + \bar{X}_1^2) >$$
$$-2\alpha\kappa\tau_p T_{norm}\bar{X}_1^2\bar{Y}_1\cos(\Delta\bar{\Phi}) \quad (A9)$$

In the case of zero detuning, this can be simplified to:
$$\kappa\tau_p > -\frac{\alpha T_{norm}\bar{X}_1^2\bar{Y}_1}{2T_{norm}(\gamma_c + \bar{X}_1^2)}, \text{when } \Delta\bar{\Phi} = 0, \quad (A10a)$$

And
$$\kappa\tau_p > \frac{\alpha T_{norm}\bar{X}_1^2\bar{Y}_1}{2T_{norm}(\gamma_c + \bar{X}_1^2)}, \text{when } \Delta\bar{\Phi} = \pi \quad (A10b)$$

For the condition $B_1B_2 - B_3 > 0$ to hold true, the real part of the complex solutions to $\lambda$ must be negative. Consequentially, this yields a second order equation for $\kappa$,
$$C_1\kappa^2 + C_2\kappa + C_3 > 0 \quad (A11)$$

where
$$C_1 = -4\cos^2(\Delta\bar{\Phi})(\bar{Y}_1 - 1) \quad (A12a)$$
$$C_2 = -2\alpha T_{norm}\cos(\Delta\bar{\Phi})\bar{X}_1^2\bar{Y}_1 \quad (A12b)$$
$$C_3 = T_{norm}[T_{norm}(\gamma_c + \bar{X}_1^2) - (\bar{Y}_1 - 1)] \times$$
$$\left(\frac{\beta\tau_{nr}}{\tau_{rad}}\bar{Y}_1 - \gamma_c\bar{Y}_1 + \gamma_c + \bar{X}_1^2\right) \quad (A12c)$$

An explicit expression describing the stable phase-locking conditions is challenging to obtain. Nevertheless, we can plot out and observe that the Hopf bifurcation boundary with $C_1\kappa^2 + C_2\kappa + C_3 = 0$, is a parabolic function, whose center and width vary with $\beta$ and $P$. Each set of parameters generates a different parabolic function, and generates either zero, one or two roots, as shown in Fig. 1(a).

REFERENCES

[1] C. Poulton *et al*., "Coherent solid-state LIDAR with silicon photonic optical phased arrays," *Opt. Lett.*, vol. 42, no. 20, 2017, pp. 4091-4094.
[2] B. Lee *et al*., "Beam combining of quantum cascade laser arrays," *Opt. Exp.,* vol. 17, no. 18, 2009, pp. 16216-16224.
[3] D. Kedar and S. Arnon, "Urban optical wireless communication networks: the main challenges and possible solutions," *IEEE Communications Magazine*, vol. 42, no. 5, 2004, pp. S2-S7.
[4] D. Kwong *et al*., "On-chip silicon optical phased array for two-dimensional beam steering." *Opt. Lett., vol.* 39, no. 4, 2014, pp. 941-944.
[5] J. K. Doylend, M. J. R. Heck, J. T. Bovington, J. D. Peters, L. A. Coldren, and J. E. Bowers. "Two-dimensional free-space beam steering with an optical phased array on silicon-on-insulator." *Opt. Exp., vol.* 19, no. 22, 2011, pp. 21595-21604.
[6] C. Henry. "Theory of the linewidth of semiconductor lasers". *IEEE J. Quantum Electron., vol. 18, no. 2,* 1982, pp. 59-64.
[7] S. Shahin, F. Vallini, F. Monifi, M. Rabinovich and Y. Fainman, "Heteroclinic dynamics of coupled semiconductor lasers with optoelectronic feedback," *Opt. Lett.,* vol. 41, no.22, 2016, pp. 5238-5241, 2016.
[8] B. Bahari, A. Ndao, F. Vallini, A. El Amili, Y. Fainman and B. Kante, "Nonreciprocal lasing in topological cavities of arbitrary geometries," *Science,* vol. 358, no. 6363, 2017, pp. 636-640.
[9] M. P. Hokmabadi, N. S. Nye, R. El-Ganainy, D. N. Christodoulides and M. Khajavikhan, "Supersymmetric laser arrays," *Science*, vol. 363, no. 6427, 2019, pp. 623-626.
[10] M. Marconi *et al*., "Mesoscopic limit cycles in coupled nanolasers," *Phys. Rev. Lett.,* vol. 124, no. 21, 2020, p. 213602
[11] Z. Jia, L. Wang, J. Zhang, Y. Zhao, C. Liu, S. Zhai, N. Zhuo, J. Liu, L. Wang and S. Liu, "Phase-locked array of quantum cascade lasers with an intracavity spatial filter," *Appl. Phys. Lett.*, vol. 111, no. 6, 2017, p. 061108.
[12] T-Y. Kao, J. L. Reno and Q. Hu, "Phase-locked laser arrays through global antenna mutual coupling," *Nature Photon.*, vol. 10, no. 8, 2016, p. 541.
[13] S. S. Wang, H. G. Winful. "Dynamics of phase-locked semiconductor laser arrays". *Appl. Phys. Lett., vol. 52, no. 21,* 1988, pp. 1774-1776.
[14] H. Winful and S. Wang, "Stability of phase locking in coupled semiconductor laser arrays," *Appl. Phys. Lett*, vol. 53, no. 20, 1988, pp. 1894-1896.
[15] M. Adams, D. Jevtics, M. J. Strain, I. Henning and A. Hurtado, "High-frequency dynamics of evanescently-coupled nanowire lasers," *Scie. Rep.*, vol. 9, no. 1, 2019, pp. 1-7.
[16] M. Adams, N.Li, B. Cemlyn, H. Susanto and I. Henning, "Effects of detuning, gain-guiding, and index antiguiding on the dynamics of two

laterally coupled semiconductor lasers," *Phys. Rev. A*, vol. 95, no. 5, 2017, p. 053869.
[17] H. Erzgraber, S. Wieczorek and B. Krauskopf, "Dynamics of two laterally coupled semiconductor lasers: Strong-and weak-coupling theory," *Phys. Rev. E*, vol. 78, no. 6, 2008, p. 066201.
[18] J. Shena, J. Hizanidis, V. Kovanis, and G. P. Tsironis. "Turbulent chimeras in large semiconductor laser arrays." *Sci. Rep.,* vol. 7, 2017, pp. 42116.
[19] D.T. Cassidy. "Spontaneous-emission factor of semiconductor diode lasers," *J. Opt. Soc. Am. B*, vol. 8, no. 4, 1991, pp. 747–752.
[20] M. Khajavikhan *et al*., "Thresholdless nanoscale coaxial lasers," *Nature*, vol. 482, no. 7384, 2012, pp. 204-207.
[21] Q. Gu, B. Slutsky, F. Vallini, J. S. Smalley, M. P. Nezhad, N. C. Frateschi and Y. Fainman, "Purcell effect in sub-wavelength semiconductor lasers," *Opt. Express*, vol. 21, 2013, pp. 15603-15617.
[22] S. H. Pan, Q. Gu, A. El Amili, F. Vallini and Y. Fainman, "Dynamic hysteresis in a coherent high-$\beta$ nanolaser," *Optica*, vol. 3, no. 11, 2016, pp. 1260-1265.
[23] C.-Y. Fang *et al.* , "Lasing action in low-resistance nanolasers based on tunnel junctions," *Opt. Lett.*, vol. 44, no. 15, 2019, pp. 3669-3672.
[24] S. Deka *et al.,* "Real-time dynamic wavelength tuning and intensity modulation of metal-clad nanolasers", *Opt. Exp.* , vol. 28, no. 19, 2020, pp. 27346-27357
[25] R. M. Ma, R.F. Oulton. "Applications of nanolasers," *Nat. Nanotech.*, vol. 14, no. 1, 2019, pp. 12-22.
[26] Q. Gu *et al.,* "Subwavelength semiconductor lasers for dense chip-scale integration," *Adv. in Optics and Photonics* 2014; 6:1-56.
[27] P. Hamel *et al.,* "Spontaneous mirror-symmetry breaking in coupled photonic-crystal nanolasers," *Nat. Photon.*, vol. 9, no. 5, pp. 311-315, 2015.
[28] E. Schlottmann *et al*., "Injection locking of quantum-dot microlasers operating in the few-photon regime." Phys, Rev. Appl., vol. 6, no. 4, 2016, pp.044023.
[29] S. Kreinberg *et al*., "Mutual coupling and synchronization of optically coupled quantum-dot micropillar lasers at ultra-low light levels." *Nat. Commun.*, vol. 10, no. 1, 2019, pp. 1-11.
[30] T. Suhr T, P. Kristensen PT, J. Mørk, "Phase-locking regimes of photonic crystal nanocavity laser arrays," Appl. Phys. Lett., vol. 99, no. 25, 2011, p. 251104.
[31] S. Deka, S. H. Pan, Q. Gu, Y.Fainman, and A. E. Amili. "Coupling in a dual metallo-dielectric nanolaser system." *Opt. Lett., vol.* 42, no. 22 2017, pp.4760-4763.
[32] L. Coldren, S. Corzine, and M. Mashanovitch. *Diode lasers and photonic integrated circuits*. Hoboken, NJ, USA: Wiley, 2012, pp. 247-329
[33] K. Yu, A. Lakhani, and M. Wu. "Subwavelength metal-optic semiconductor nanopatch lasers." Optics express 18, no. 9 (2010): 8790-8799.
[34] Y. Halioua *et al*., "Hybrid III-V semiconductor/silicon nanolaser." *Opt. Express, vol.* 19, no. 10 (2011): 9221-9231.
[35] Y. Zhao, C. Qian, K. Qiu, Y. Gao and X. Xu, "Ultrafast optical switching using photonic molecules in photonic crystal waveguides," *Opt. Exp.,* vol. 23, no. 7, pp. 9211-9220, 2015.
[36] K. A. Atlasov, K. F. Karlsson, A. Rudra, B. Dwir and E. Kapon, "Wavelength and loss splitting in directly coupled photonic-crystal defect microcavities," *Opt. Exp.,* vol. 16, no. 20, pp. 16255-16264, 2008.
[37] S. Haddadi *et al*., "Photonic molecules: tailoring the coupling strength and sign." *Opt. Exp*., vol.22, no. 10, 2014, pp. 12359-12368.
[38] Z. Gao, D. Siriani, and K. D. Choquette. "Coupling coefficient in antiguided coupling: magnitude and sign control." *JOSA B,* vol. 35, no. 2 (2018): 417-422.
[39] B. Ermentrout, *Simulating, analyzing, and animating dynamical systems: a guide to XPPAUT for researchers and students*. Philadelphia , PA, USA: Society for Industrial and Applied Mathematics, 2002.
[40] P. Antonik, Paul, M.C. Wicks, H. D. Griffiths, and C.J. Baker. "Frequency diverse array radars." in *IEEE Conference on Radar*, 2006, p. 3-pp.
[41] A. Yao, W. Wu, and D. Fang. "Frequency diverse array antenna using time-modulated optimized frequency offset to obtain time-invariant spatial fine focusing beampattern." *IEEE Trans. on Antennas and Propag.*, vol. 64, no. 10, 2016, pp. 4434-4446.
[42] Y. Xu, X. Shi, J. Xu, and P. Li. "Range-angle-dependent beamforming of pulsed frequency diverse array." *IEEE Trans. on Antennas and Propag.*, vol. 63, no. 7, 2015, pp. 3262-3267.
[43] G. Crosnier *et al.,* "Hybrid indium phosphide-on-silicon nanolaser diode." *Nat. Photon.,* vol. 11, no. 5, 2017, pp. 297-300.
[44] S. Pan, S. Deka, A. El Amili, Q. Gu, and Y. Fainman. "Nanolasers: second-order intensity correlation, direct modulation and electromagnetic isolation in array architectures." *Prog. in Quantum Electron.*, vol. 59, 2018, pp. 1-18.
[45] L. Liu *et al.,* "An ultra-small, low-power, all-optical flip-flop memory on a silicon chip." *Nat. Photon.,* vol. 4, no. 3, 2010, pp. 182-187.
[46] P. McMahon *et al.*, "A fully programmable 100-spin coherent Ising machine with all-to-all connections." *Science*, vol. 354, no. 631, 2016 , pp. 614-617.